\documentclass[12pt]{article}

\usepackage{scicite}

\usepackage{times}

\usepackage{graphicx}
\usepackage{lscape}

\newenvironment{sciabstract}{%
\begin{quote} \bf}
{\end{quote}}

\title{Poverty levels, societal and individual heterogeneities explain the SARS-CoV-2 pandemic growth in Latin America}

\author
{Jos\'e M. Ponciano$^{1\ast}$, Juan A. Ponciano$^{2\ast}$, Juan P.  Gomez$^{3}$,\\ 
Robert D. Holt$^1$, and Jason K. Blackburn$^{4,5}$.\\
\\
\normalsize{$^{1}$Department of Biology, University of Florida, Gainesville, FL 32611, USA.}\\
\normalsize{$^{2}$Instituto de Investigaci\'on en Ciencias F\'isicas y Matem\'aticas,}\\ 
\normalsize{Universidad de San Carlos de Guatemala,  Guatemala.}\\
\normalsize{$^{3}$Departamento de Qu\'imica y Biolog\'ia,}\\ 
\normalsize{Universidad del Norte, Barranquilla, Colombia.}\\
\normalsize{$^{4}$Spatial Epidemiology and Ecology Research Laboratory,}\\
\normalsize{Department of Geography,  University of Florida, Gainesville, FL 32611, USA,}\\
\normalsize{$^{5}$ Emerging Pathogens Institute, University of Florida, Gainesville, FL 32611, USA.}\\
\\
\normalsize{$^\ast$Correspondence to:  josemi@ufl.edu (J.M.P.); japonciano@ecfm.usac.edu.gt (J.A.P.)}
}

\date{}

\begin{document} 


\baselineskip24pt


\maketitle


\begin{sciabstract}
Latin America is experiencing severe impacts of the SARS-CoV-2 pandemic, but poverty and weak public health institutions hamper gathering the kind of refined data needed to inform classical SEIR models of epidemics.  We present an alternative approach that draws on advances in statistical ecology and conservation biology to enhance the value of sparse data in projecting and ameliorating epidemics.  Our approach, leading to what we call a Stochastic Epidemic Gompertz model, with few parameters can flexibly incorporate heterogeneity in transmission within populations and across time. We demonstrate that poverty has a large impact on the course of the pandemic, across fourteen Latin American countries, and show how our approach provides flexible, time-varying projections of disease risk that can be used to refine public health strategies.
\end{sciabstract}
\noindent
{\bf One Sentence Summary:} The growth modality of SARS-CoV-2 among Latin-American countries is well-explained by poverty differences, individual and temporal heterogeneities.

\section*{}
``$\ldots$Major global crises$\ldots$ demand cooperative global responses that do not leave out the poor. Once SARS-CoV-2 is under control, the world cannot return to business as usual'' von Braun et al. \cite{von2020} concluded in a recent editorial commentary in Science entitled ``The moment to see the poor.'' As the recent flurry of research on the SARS-CoV-2 pandemic shows, the languages of mathematics, statistics and computer science are essential instruments for grappling with the uncertain course of the pandemic. Joel Cohen's \cite{cohen2004mathematics} remark almost twenty years ago that ``mathematics is biology's next microscope, only better'' has never been more salient. Deterministic, epidemiological models of the SEIR type (Susceptible $S(t)$, Exposed $E(t)$, Infected $I(t)$ and Recovered $R(t)$) have long enabled an in-depth exploration of infectious disease processes \cite{bauch2005dynamically,earn2000simple,chowell2016mathematical,lipsitch2003transmission}, and provide a framework of fundamental principles to manage infectious diseases \cite{chowell2004basic,gao2011seasonality,lekone2006statistical,prem2020effect,wearing2005appropriate,chowell2003sars}   including the SARS-CoV-2 virus pandemic \cite{prem2020effect,li2020substantial,read2020novel,yang2020modified}.  While these models are useful, parameterizing them for a novel viral pandemic with limited diagnostics and systematic data collection approaches can be challenging.  We address this need \cite{von2020} with a complementary, multi-model approach \cite{taper2008model} that incorporates social, individual and temporal heterogeneities.  

Approaches employing simple yet biologically sound models with few parameters are particularly needed in regions like Latin America where sound strategies to collect public health data are seldom employed. Here we show that a multi-model, multi-stages modeling approach helps elucidate \textit{i}) early epidemic growth in fourteen Latin-American countries \textit{ii}) the role of poverty in shaping the growth rate of the number of cases and \textit{iii}) the probability that the number of cases of SARS-CoV-2 exceeds any given amount within arbitrarily defined small windows of time, starting from the present. 

Characterizing complex epidemiological processes depends on the adequate formulation of proper probabilistic models of governing processes. Survival, extinction and growth of natural populations --including infectious diseases-- are inherently stochastic \cite{dennis1991estimation}. At its core, the problem of modeling the growth of the accumulation of SARS-CoV-2 cases is a stochastic population dynamics problem begging for a full characterization of the probabilities of the possible trends. Computer intensive methods to fit biologically sound stochastic models to the growth of cases permit estimation of such probabilities. These methods notably allow a process-based estimation of the time-varying probability that within a given window of time, the number of cases will exceed any given number grounded in the dynamics of the infection process. Mathematically, this problem is analogous to the conservation biology goal of estimating extinction risk for endangered populations.

The application of stochastic processes to the estimation of extinction risks and average times until an event of interest happens has enriched conservation biology \cite{staples2005risk}, wildlife management \cite{dennis2006estimating,ferguson2014predicting} and evolutionary microbial population dynamics \cite{ponciano2007population}. In conservation biology, Population Viability Analyses (PVA) were transformed  by stochastic processes used to predict populations' growth and extinction \cite{dennis1991estimation}, despite complexities of the target species' life-cycle \cite{taper2008model}. Ideally, a population dynamics model is relatively simple yet retains essential features of demographic and environmental stochasticities \cite{dennis1991estimation,allen2017primer} as well as robustness in the face of other sampling complexities \cite{taper2008model}. These features are particularly important given the paucity of complete data/information.  The analysis of SARS-CoV-2 time series data from countries with scant public health resources is not unlike the study of endangered species time series data -- both seek to use whatever information is available to distill the general principles governing data fluctuations in order to then make informed projections and assess ``what-if'' scenarios.   

It is now recognized that long-term estimates of persistence probabilities are of little use because of the ever-changing conditions of population growth \cite{staples2005risk}. Viable Population Monitoring (VPM) aims at estimating and updating persistence probabilities over short time horizons using fresh data  \cite{staples2005risk}. Such stepwise, continuously updated, estimates of short-term persistence probabilities are more practical and actionable than long-term projections. We draw on prior work in conservation biology, population dynamics and epidemiological theory to complement the current suite of deterministic epidemiological models, characterize the role of urban poverty in shaping the region's SARS-CoV-2 epidemics, and develop a methodology to generate short (5-15 days), sequentially updatable, process-based forecasts.

\paragraph*{Early epidemics : patterns and processes}
Characterizing the early phase of the epidemic growth profiles in Latin America (Fig. 1A-D) provides a first step towards understanding the transmission dynamics of SARS-CoV-2 in the region. Unconstrained growth is often a fair assumption at the outset of an emerging disease, so growth of the number of cases $n(t)$ over time are properly described by an exponential growth model wherein the per capita rate of change in total case numbers is the constant intrinsic growth rate $r$: 
\begin{equation}
\frac{1}{n(t)}\frac{ dn(t)}{dt}=r. 
\end{equation}
Accordingly, the per capita contribution to the growth of the epidemic is unaffected by the total number of cases. Early in an epidemic, $r$ has been used to estimate the basic reproduction number $R_0$ using approximate relations \cite{lipsitch2003transmission,heffernan2005perspectives} but their accuracy is model-dependent.  Epidemic growth is frequently limited by many factors, including reactive behavior changes or spatially constrained contact structures.  However diverse, these factors tend to act as ``density-dependent'' processes with slower growth patterns \cite{chowell2016mathematical}. Chowell et al. \cite{chowell2016mathematical} note that different compartmental models all lead to an early sub-exponential growth. A key factor is inhomogeneous mixing in contact rates, formalized as a non-linear contact rate function of the type $dI(t)/dt=\beta S(t)I(t)^{1-b}/N(t),\; 0<b<1$. The resulting sub-exponential growth describes epidemiological data reported in the literature for many important epidemics \cite{chowell2016mathematical}. Importantly, this model encapsulates, albeit phenomenologically, the outcome of lower level mechanistic models \cite{chowell2016mathematical} where clustering, structuring, other forms of heterogeneity and the mean and variance of the contact distribution are ultimately responsible for the degree of sub-exponentiality.  

The key observation triggering our research was that early sub-exponential growth in fourteen Latin-American countries could be clearly divided into four distinct growth profiles (Fig. 1A-D) according to its approach to pure exponential growth. We then hypothesized that such differences in the degree of sub-exponential growth could then be explained by differences in poverty, which we expect to modulate the distribution of contacts and possibly other mechanisms \cite{chowell2016mathematical}.      

\begin{figure}
\includegraphics[width=\textwidth]{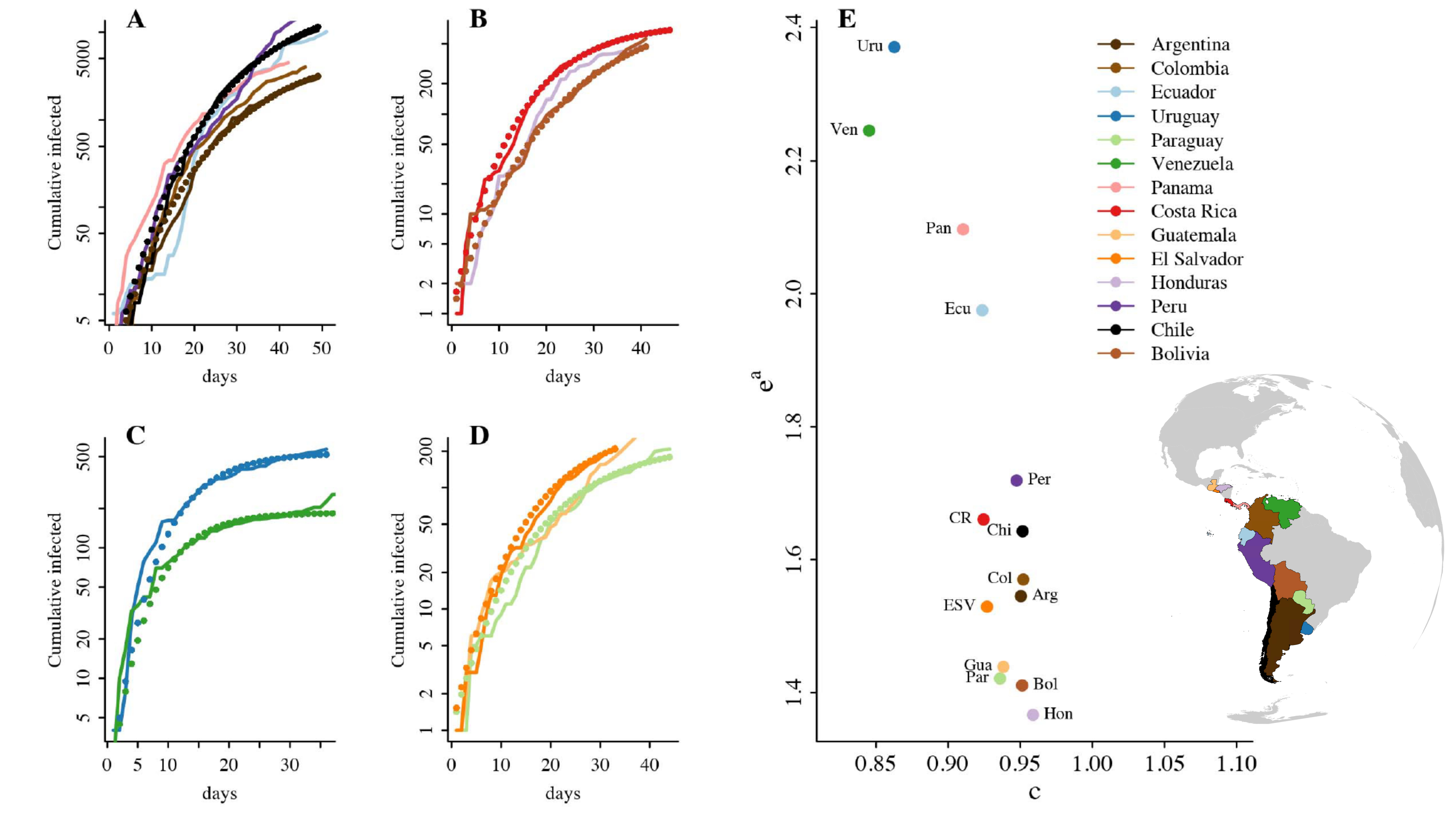}
\caption{SARS-CoV-2 cumulative cases for 14 Latin-American countries (solid colored lines). ({\bf A-D}) Deterministic fit of the EG model (filled circles) with a Poisson likelihood is shown for fourteen countries grouped according to growth modality. Data includes the counts from the beginning of the epidemic until April $20^{th}$. ({\bf E}) Estimates of the EG model parameters plotted against each other.}
\end{figure}

Compartmental, multivariate SEIR models are an every-day indispensable tool to obtain conceptual and practical insights. But these models are data-hungry and statistically costly \cite{campbell2014anova} due to their (potentially) large number of parameters. The ratio of data to the number of parameters needing estimation remains a challenge in Latin America due to a lack of data gathering infrastructure, cohesive contact tracing plans, and testing and monitoring capabilities. Despite these challenges, we show that in twelve out of these fourteen countries, including some form of heterogeneity (\textit{e.g.} structuring according to age or poverty or, due to inhomogeneous mixing of susceptible and infected individuals \cite{hethcote2000mathematics}) improves model fits compared with the classic SEIR model mostly used to date (Table 1). First, we fitted the following SEIR type deterministic model variants: a classic SEIR model with and without non-homogeneous mixing, and a structured population SEIR model with and without non-homogeneous mixing with structuring according to poverty and age class.  The two forms of demographic structuring and the non-homogeneous mixing were included to assess the effect of different sources of heterogeneity.   For all 14 countries we modeled the start of the epidemic as resulting from imported cases and included the effect of each nation's airport-closing decision (see Table 1 and Supplementary Material).     

Next, our analyses focused on the development of actionable theory-grounded univariate models requiring fewer parameters. These models incorporate i) different degrees of heterogeneity among hosts in pathogen transmission and ii) variability in the dynamics due to overall poverty levels. The overall time series variability is decomposed into sampling error and two forms of process error: demographic and environmental variabilities \cite{dennis1991estimation,ferguson2015evidence,ponciano2009hierarchical}.

\paragraph*{The early epidemic involves demographic stochasticity}
To model the dynamics of initial infection, we used a stochastic pure birth process -- a continuous time and discrete states Markov Process used in various epidemiological contexts \cite{allen2017primer,allen2010introduction,allen2012extinction}. The type of variability displayed here, demographic stochasticity,  \cite{dennis1991estimation,ferguson2014predicting,allen2017primer,allen2010introduction}  represents chance variation in infection due to heterogeneities in individual contact rates. This type of process variability looms large at low numbers of infected individuals \cite{dennis1991estimation}.  Let $N(t)$ be the random number of accumulated cases in a country at time $t$ and $p_{n}(t)=P(N(t)=n(t))$ the probability of observing $n(t)$ cases at time $t$.  We introduced an inhomogeneous contact rate function to obtain a form of the birth rate $\lambda_n$ that leads to sub-exponential growth early in an epidemic. How to incorporate heterogeneity into a univariate model is detailed next. Either analytically or numerically we calculate $p_{n}(t)$ and use it to compute the probability that the process exceeds a given threshold $n_c$ within a pre-determined future time interval (Fig. S2), as well as first passage times (fpts) defined as $T(n_c):=\rm{min}(\tau >0: N(\tau) \geq n_c |N_0=n_0)$.

Pure birth processes become quasi-deterministic at large population sizes (cases), but process variability remains important.  When case numbers are large, deviations from deterministic predictions can emerge from temporal (or spatial) variation in the transmission rate, known as environmental stochasticity \cite{dennis1991estimation,ferguson2014predicting,allen2017primer} (in addition to observation error 
\cite{dennis2006estimating}).  Spatial heterogeneities may be determined by socio-economic factors.    We build a hierarchical model of the accumulation of the total number of cases including poverty and heterogeneity in transmission to jointly fit to 14 Latin-American countries (Fig. 1A-D). We then extend this model formulation to include environmental variability and use it to formulate practical risk assessment tools for each country.

\paragraph*{Environmental stochasticity, heterogeneity and the role of poverty}
Our theoretical derivation begins with the specification of the stochastic number of successful transmission encounters that each one of the total cases up to day $t$, $n(t)$, generates from one day to the next.  The expected number of cases in consecutive days is obtained as follows:  initially, each case $i$ incurs in a random number of contacts $X_i, \; i=1,2,\cdots,n(t)$ with susceptible individuals. Out of these, a random number result in a successful transmission encounter. If on average these individuals contact $m$ susceptible individuals, and if $p(t)$ is the probability that one such contact results in a successful transmission,  then the average number of cases for the following day is $n(t) m p(t)$. Recovery is included in the term $p(t)$ (see below).  A similar framework is used in epidemiology \cite{lloyd2005superspreading} yet here we refrain from adopting a Poisson or negative binomial distribution for any of these quantities.  This level of generality has two consequences.  First, a moments-based decomposition of the process' variance of the growth rate  (Supplementary Material) given demographic, environmental and individual variabilities and any combination thereof is obtained with the added benefit of not being tied to particular scalings.  Parametric models contain implicit relationships, such as variance scalings, and adopting any particular model has unforeseen consequences for the estimation of risk quantities \cite{ponciano2018parametric,ferguson2015evidence}. Our analysis complements previous formulations in both epidemiology and population dynamics \cite{ferguson2014predicting,lloyd2005superspreading} and reflects what one might call the ``Cohen principle,'' which demonstrates that distributions of abundance do not provide a shortcut to understanding the mechanisms that generate those distributions. Each requires analysis \cite{cohen2020every}. Our minimal assumptions approach has direct, practical consequences: it re-directs the inferential focus towards biologically relevant variance forms of $N(t)$. These are key to accurately estimate population risks \cite{ferguson2015evidence}.

Second, this framework is amenable to multiple parameterizations regarding infection dynamics, for example, by assuming inhomogeneity in $p(t)$. Using Eq. (6) in \cite{chowell2016mathematical}  (an expression for inhomogeneous mixing during infection) we let $p(t)=I(t)^{-b},~0<b<1$,  where $I(t)$ is the number of infected that day.  If $\gamma$ is the per-day recovery probability, then $I(t)^{-b}=(n(t)-n(t)\gamma)^{-b}$ so that the average one-step change in the number of cases is:   
\begin{equation}
n(t+1)= \frac{m}{(1-\gamma)^b} n(t)^{1-b}.
\end{equation}
Letting $m'=m/(1-\gamma)^b \equiv e^a, ~a>0$  Eq. (2) is seen to be a discrete time Gompertz model. We call this model the Epidemic Gompertz (EG) model.  Because the solution of the discrete Gompertz map matches the solution of the ODE Gompertz equation in continuous time \cite{ponciano2018parametric} $dN(t)/dt= \theta N(t) ( \log \kappa - \log N(t))) $,  there is a one-to-one relation between our EG model and the ODE model parameters. In particular, $c=1-b=e^{-\theta}$ and $a=b \log \kappa$, where $\kappa$ is the equilibrium total number of infected and $\theta$ is the speed of equilibration. Note that  $c$ is a simple transformation of the speed of equilibration of the continuous time Gompertz equation.  When $c$ is 1, the process grows without bounds.  In the epidemiological context, that means that  $b=0$ thus implying homogeneous mixing. The closer the value of $c$ is to 1, the closer the epidemic model is to homogenous mixing and the closer the growth is to pure exponential growth initially (see Fig 1E for estimates for $e^a$ and $c$ for Latin America). Conversely, the smaller the value of $c$ (\textit{i.e.}, the higher the value of $b$), the stronger the degree of inhomogeneity.   

To characterize the role of poverty, we tested a statistical hierarchical model where the inhomogeneity parameter $c$ in the EG model (Eq. 2) depends on urban poverty as reported by the United Nations Economic Commission for Latin -America and the Caribbean (UN-ECLAC, see Supplementary Material) against a null Gompertz model where such inhomogeneity was unconstrained and not related to poverty. Specifically, the model postulates that each countries' inhomogeneity parameter is drawn from a random effect whose mean depicts a non-linear relationship between $c=1-b$ and urban poverty.  Lognormal sampling error is added in both the null and the alternative model. A likelihood ratio test for hierarchical models rejected the null model (p-value=0.036). The joint, multi-countries hierarchical model with poverty as a covariate was about $8.92$ times more likely to explain the ensemble of time series data than a model without the poverty covariate as given by the likelihood ratio test. Higher urban poverty index yields on average (thick black line in Fig. 2A) values of $c$ closer to 1 (\textit{i.e.} higher homogeneity in contact rates), which in turn implies that the accumulation of cases is closer to exponential growth (Fig. 2A). Indeed, greater social homogeneity in poorer countries means a higher proportion of the population lives in poverty.  

Having demonstrated the significance of the urban poverty covariate, we repeated the estimation without Venezuela because it was the only country for which we had difficulties cross-checking the data from multiple sources. Transparency in data reporting is paramount for scientific inference. This time, besides the random poverty-driven effect, we postulated that the growth of the epidemic was dominated by environmental stochasticity.  In this new model, Eq. (2) in the natural logarithm scale is the mean of a Markovian transition probability distribution. We called this model the Stochastic Epidemic Gompertz (SEG) model.  Using the SEG model, we obtained a tighter relationship (Fig. 2B) between poverty and the inhomogeneity parameter, despite the added layer of randomness.  Just as before, higher urban poverty index yields on average (thick black line in Fig. 2B) values of $c$ closer to 1 (\textit{i.e.}, higher homogeneity in contact rates). Because the SEG is a model  with environmental stochasticity, it allows accommodating a large mean number of contacts per individual while keeping the variance of the ``offspring'' distribution null (\textit{i.e.}, no demographic stochasticity, only environmental noise). In network theory models, this would amount to specifying a distribution of the number of contacts that has a large mean but a very small (if not null) variance.  Our model construction process focuses on the specification of the nature of the variability (see Supplementary Material), and hence can readily accommodate many other mean to variance relationships besides the one implied by the SEG model. The point is, the contact and infection processes \cite{ponciano2011first}, not the distributional assumptions, take a central stage as in \cite{lloyd2005superspreading}.  The Kalman Filter (KF) applied to our poverty fit yielded a joint, time-dependent, effective reproduction number ($R_t$) predictions for the thirteen countries plotted in figure 2B (see Fig. 3 and Supplementary Material for the $R_t$ approximation).  In all cases, $R_t$ declines but remains above one. Notably, poorer countries tend to suffer higher effective reproduction numbers.

 \begin{figure}
\includegraphics[width=\textwidth]{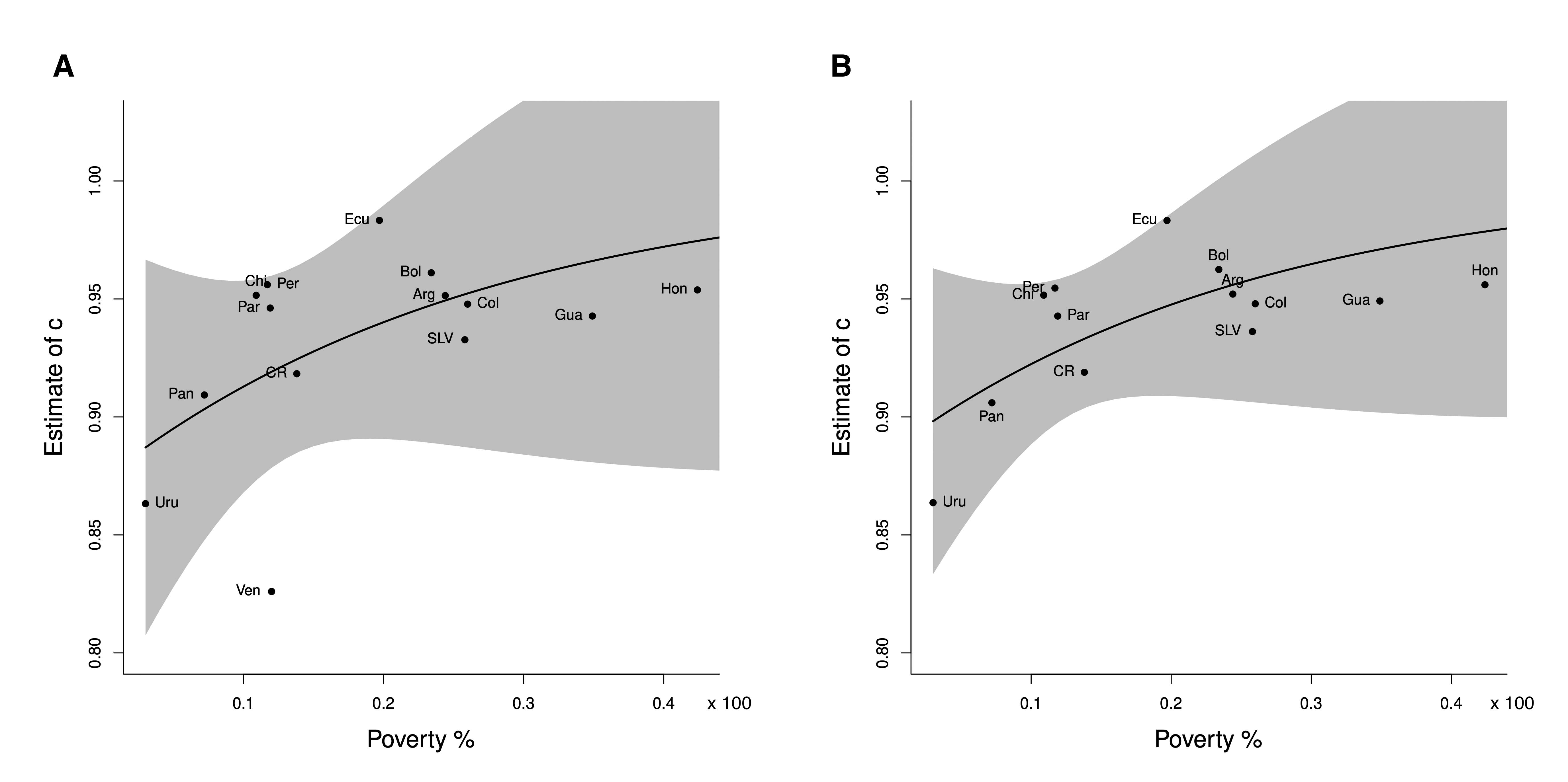}
\caption{Urban poverty vs. the inhomogeneity parameter $c$ as predicted by ({\bf A)} a hierarchical model where $c$ results from a random poverty effect whose mean (black line trend) is shaped by poverty and the growth dynamics is according to the EG model and ({\bf B}) the SEG model excluding the data from Venezuela, where now, besides the random effect link to poverty, the growth of the number of cases incorporates environmental stochasticity.}
\end{figure}

\begin{figure}
\includegraphics[width=\textwidth]{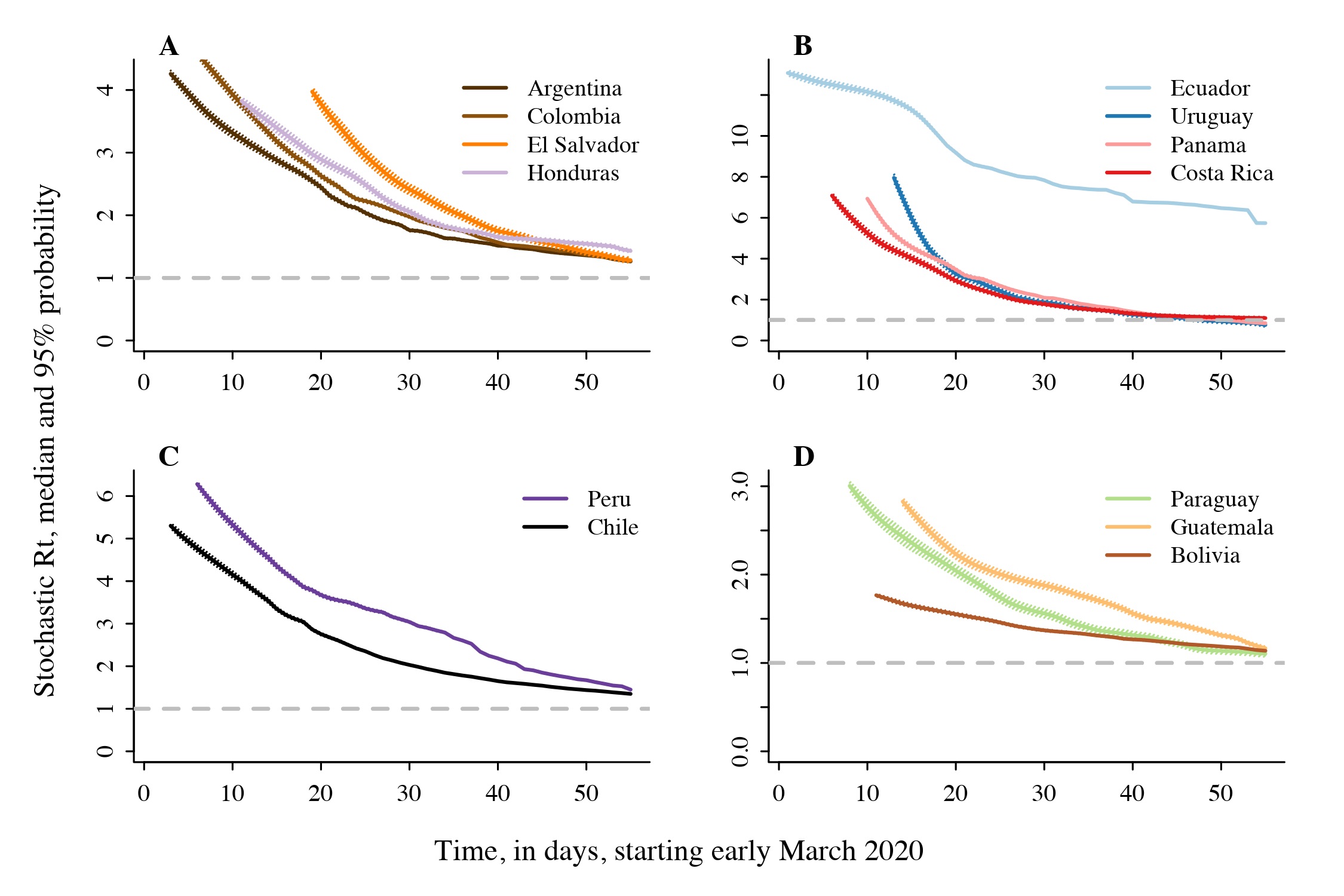}
\caption{Estimated time-dependent effective reproduction number $R_t$ for thirteen Latin-American countries grouped according to decay pattern from the start of the epidemic in each country in early March. Ecuador started the earliest on March 3$^{\rm{rd}}$ 2020.  The horizontal dashed line marks the threshold of $R_t=1$. Marked in bold are the most probable paths along with the 95\% most probable path range (hashed lines).}
\end{figure}

To illustrate the conceptual and practical benefits of complementing SEIR modeling with our SEG models we conducted two numerical experiments.  First, we compared the predictive qualities of our SEG modeling approach with the deterministic predictions of the best SEIR model variant fit in each country (Table 1 and Supplementary Material).  Second, we showed how the VPM concept  \cite{staples2005risk} can be used to build an accurate, Risk Prediction Monitoring (RPM) tool aimed at serially updating each day, the probability of recording an increase in the number of cases by any given, arbitrarily defined amount.  The trend in these serially computed probabilities can be used as a reliable indicator of whether a country is on course to control the epidemic over short forecasting horizons.

\paragraph*{Environmental stochasticity vs. SEIR models}
To compare the SEG model with the SEIR models, for each country we fitted both deterministic and stochastic models from the beginning of the epidemic until April $26^{\rm{th}}$.  Then we projected each model for sixteen days.  In the first case the projections consisted of the deterministic solution of the best-fitting ODE model for four countries (Fig. 4). Similar results for other countries are in the Supplementary Material.  For our SEG model, computing these projections amounted to simulating 50000 trajectories of 16 days using its maximum likelihood estimates.  We then plotted for the same four countries the most probable path along with the inter-quartile (IQ) range of these 50000 simulated paths (Fig. 4). This comparison (analogous to stochastic forecasting of hurricane paths) clearly shows that the hierarchical model approach is at least as good or better than the deterministic predictions (Fig. 4, Table 1).  In every case, the future observations (data towards the end of the observed time series, not part of the fitting procedure but retained for testing) are as good as (for Colombia) or closer to the most probable path than they are to the deterministic predictions. The epidemic forecasting using the SEG model thus appears more reliable for longer forecasts than the deterministic solutions. This property of the SEG model could be particularly useful in the face of sudden changes of some kind (social, political, public health policy, and so forth) in the context within which the epidemic is developing.  

\begin{figure}
\includegraphics[width=\textwidth]{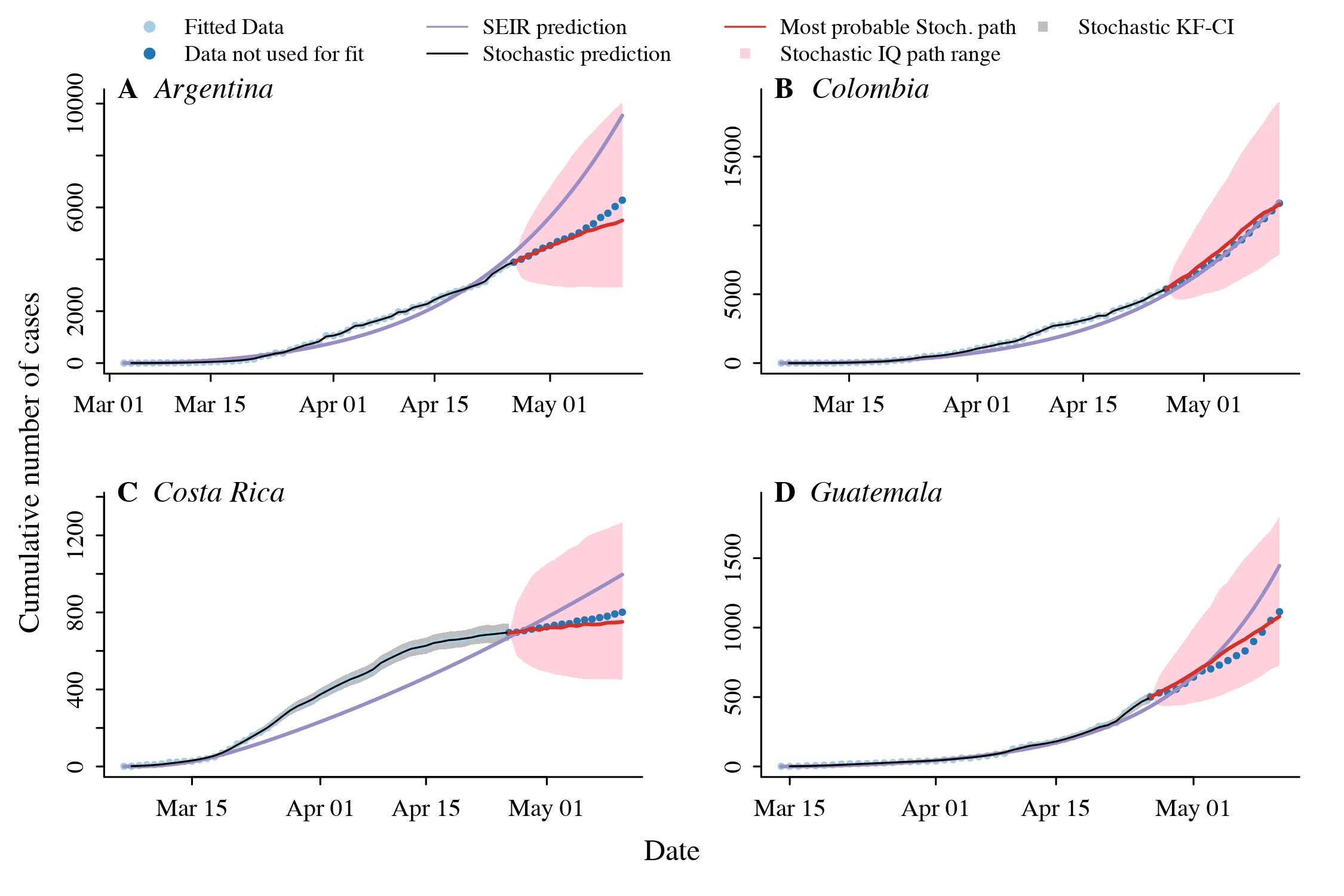}
\caption{Stochastic vs. deterministic forecasts for four countries using data from the beginning of the epidemic until April $26^{th}$ 2020. Sixteen ``future observations'' not used for the fit act as a comparison benchmark between our stochastic most probable path (thick red line) and the deterministic predictions (thick purple line). The IQ range of plausible paths (pink bands) is also displayed. Similar plots for all the countries are displayed in the Supplementary Material.}
\end{figure}

\paragraph*{Risk projections and future waves}
We developed an RPM tool that mirrors conservation biology approaches \cite{staples2005risk} to serially update the quasi-extinction probabilities with every increase of the length of the time series of population abundances.  Applied to SARS-CoV-2, this process involves using the past records of the cumulative number of cases to estimate the SEG model parameters and then use these to predict for the near future ($\tau=5$ or $10$ days) the probability that the number of cases will rise above a given critical threshold $n_{crit}$, $p_{n_{crit} } (n(t),\tau)=Pr(N(t+\tau)\geq n_{crit}|N(t)=n(t) )$. With every passing day $t'$ , the estimate of $p_{n_{crit}}(n(t'),\tau )$  is updated.  The resulting $p_{n_{crit}} (n(t'),\tau) $  trend can be used to diagnose the near future risk of an increase by any amount of the number of cases. This risk can then be propagated to compute the chances of needing the same number of extra intensive care units (ICUs), or a proportion of these according to the expected proportion of severe cases.  The trend for Costa Rica is illustrated in Fig. 5. There, the probability of a spike of 50 cases or more declined over time, indicating the dwindling of the epidemic. This approach is applicable to time series of deaths for studies aiming at assessing trends in mortality risks.

\begin{figure}
\includegraphics[width=\textwidth]{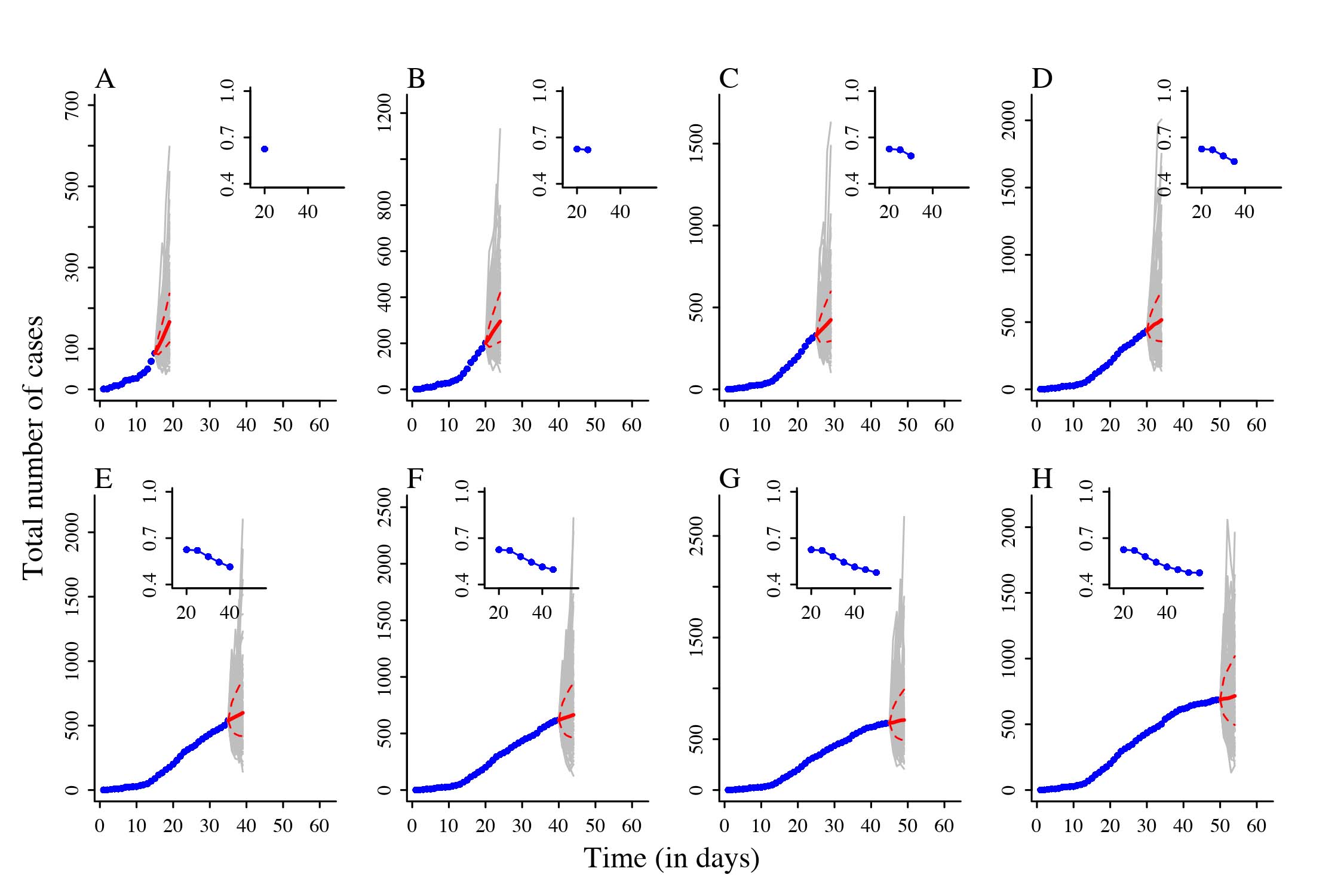}
\caption{An example of Risk Prediction Monitoring (RPM) using the Stochastic Epidemic Gompertz model fit (Eq. 2 in the main text)  for the time series of cases from Costa Rica. Inset plots show the serially updating of the risk (probability) of the number of cases increasing above 50 from the last observed count. }
\end{figure}

Finally, the advantage of introducing non-standard contact rate functions including heterogeneity is that they include non-linear phenomena like bi-stability, which, in the face of environmental stochasticity, explain how can a second wave arises without needing a phenomenological environmental forcing function (see Fig. 6 in \cite{ponciano2011first}).

\paragraph*{Conclusions}
We have used a multi-model approach to better understand and predict the SARS-CoV-2 dynamics in fourteen countries of Latin America.  Where possible, we incorporated inhomogeneous mixing and/or the effect of urban poverty.  The models we have examined and compared included compartmental SEIR models and models with demographic stochasticity and environmental noise with added sampling error and a poverty random effect (the SEG model). We used the SEG model to illustrate how with only time series of cumulative SARS-CoV-2 cases, countries with scant public health resources can make practical and conceptual advances in understanding the pandemic and forecasting its effects. 
  
Our SEG model is one of a suite aimed at decomposing the contributions from demographic, environmental and individual heterogeneities in observed time series of data. Accurately accounting for the factors shaping the variance of the growth rate of $N(t)$ yields better forecasts (Fig. 4) \cite{ferguson2014predicting,ferguson2015evidence} and the power to estimate short term trends of the risk of epidemic growth (Fig 5). These projections along with our finding that urban poverty shapes the region-wide dynamics of the SARS-CoV-2 pandemic in Latin America highlight the potential power of our approach. 

Developing reliable tools to better understand and predict complex epidemiological processes in poor countries depends on mathematical and statistical approaches attuned to a multiplicity of realities. Yes, more data are needed and always will be. Mathematical ``microscopes'' tracking the complexities of human behavior are also needed.  Yet, here we show that fundamental ecological principles can illuminate the uncertain fate of countries in need.  And this using only the most readily available source of information in these countries:  the reported time series of the total number of cases up to any given day.

\clearpage



\begin{landscape}
\begin{table}[h!]
\scriptsize

\caption{SEIR model variants parameter estimates and Root Mean Squared Errors for these models and for the SEG model applied to new data predictions (see Fig. 4). See Supplementary Materials for parameter definitions.}
\begin{tabular}{lllllllllllllllll}
\hline

     &  &      &       &     &      &     &      &       &        &        &    &    &min    &     RMSE   &   RMSE &                 \\

{\bf Country}     & $\alpha_i$  & $\alpha_r$     &  $\theta$     &  $w$    &  $\beta$    & $\sigma$     & $\gamma$     & $\mu$      & $\mu_1$       &$\mu_2$        & $N $   & {\bf BIC}   & $(\Delta BIC)$ &  SEIR&  SEG & Best   Model                \\

\hline
Argentina   & 0.88                     & 0.31 & 13.4  & 0.08 & 0.27 & 0.19 & 0.13 &       & 0.0113 & 0.0019 & 165 & 4745  & 444      & 1927           & 302        & Structured   Poverty,              \\
    &  &      &       &     &      &     &      &       &        &        &    &    &  &      &   &            Non - homogeneous mixing.       \\
Colombia    & 0.18                     & 0.04 & 60.3  & 0.08 & 0.25 & 0.19 & 0.13 &       & 0.0020 & 0.0004 & 156 & 5295  & 429      & 186            & 413        & Age   Structured, Fixed Mortality,  \\
    &  &      &       &     &      &     &      &       &        &        &    &    &  &      &   &            Non - homogeneous mixing       \\
Ecuador     & 1.00                     & 0.16 & 0.2   & 0.48 & 2.60 & 0.19 & 0.08 & 0.007 &        &        & 171 & 15632 & 502      & 5682           & 29314      & Simple,   Free Mortality,         \\
    &  &      &       &     &      &     &      &       &        &        &    &    &  &      &   &            Non - homogeneous mixing       \\
Uruguay     & 0.77                     & 0.23 & 153.3 &      & 0.08 & 0.19 & 0.08 & 0.001 &        &        & 135 & 4070  & 5        & 200            & 154        & Simple,   Fixed Mortality,             \\
    &  &      &       &     &      &     &      &       &        &        &    &    &  &      &   &            Homogeneous mixing      \\
Paraguay    & 0.17                     & 0.04 & 9     & 0.13 & 0.23 & 0.19 & 0.15 &       & 0.0020 & 0.0004 & 150 & 982   & 30       & 170            & 132        & Age   Structured, Fixed Mortality, \\
    &  &      &       &     &      &     &      &       &        &        &    &    &  &      &   &            Non - homogeneous mixing      \\
Venezuela   & 1.00                     & 0.20 & 86.4  &      & 0.07 & 0.19 & 0.11 &       & 0.0000 & 0.0058 & 132 & 1356  & 19       & 58             & NA         & Poverty   structured,                  \\
    &  &      &       &     &      &     &      &       &        &        &    &    &  &      &   &            Homogeneous mixing       \\
Panama      & 0.99                     & 0.04 & 37.3  &      & 0.22 & 0.19 & 0.12 & 0.004 &        &        & 144 & 3465  & 83       & 3435           & 204        & Simple,   Free Mortality,           \\
    &  &      &       &     &      &     &      &       &        &        &    &    &  &      &   &            homogeneous mixing       \\
Costa Rica  & 1.00                     & 0.15 & 13.9  & 0.18 & 0.14 & 0.19 & 0.08 &       & 0.0020 & 0.0004 & 156 & 2713  & 298      & 110            & 17         & Age   Structured, Fixed Mortality,
\\
    &  &      &       &     &      &     &      &       &        &        &    &    &  &      &   &            Homogeneous mixing.       \\
Guatemala   & 1.00                     & 0.12 & 6.5   &      & 0.28 & 0.19 & 0.18 &       & 0.0036 & 0.0077 & 132 & 731   & 14       & 165            & 48         & Structured   Poverty,                 \\
    &  &      &       &     &      &     &      &       &        &        &    &    &  &      &   &            Homogeneous mixing       \\
El Salvador & 0.37                     & 0.06 & 11.2  & 0.19 & 0.19 & 0.19 & 0.12 &       & 0.0020 & 0.0004 & 117 & 704   & 277      & 270            & 142        & Age   Structured, Fixed Mortality,\\
    &  &      &       &     &      &     &      &       &        &        &    &    &  &      &   &            Non - homogeneous mixing       \\
Honduras    & 0.13                     & 0.01 & 78.5  & 0.12 & 0.19 & 0.19 & 0.10 &       & 0.0020 & 0.0004 & 141 & 1410  & 12       & 567            & 284        & Age   Structured, Fixed Mortality \\
    &  &      &       &     &      &     &      &       &        &        &    &    &  &      &   &            Non - homogeneous mixing      \\
Peru        & 0.99                     & 0.84 & 16.8  & 0.08 & 0.22 & 0.19 & 0.08 & 0.005 &        &        & 156 & 19751 & 1209     & 266000         & 3765       & Simple,   Free Mortality,        \\
    &  &      &       &     &      &     &      &       &        &        &    &    &  &      &   &            Non - homogeneous mixing.      \\
Chile       & 0.69                     & 0.40 & 22.6  & 0.12 & 0.32 & 0.19 & 0.11 &       & 0.0020 & 0.0004 & 165 & 22117 & 2746     & 1977           & 868        & Age   Structured, Fixed Mortality,\\
    &  &      &       &     &      &     &      &       &        &        &    &    &  &      &   &            Non - homogeneous mixing       \\
Bolivia     & 1.00                     & 0.05 & 2.5   & 0.26 & 0.45 & 0.19 & 0.34 & 0.025 &        &        & 141 & 1035  & 138      & 267            & 182        & Simple,   Free Mortality,   \\
    &  &      &       &     &      &     &      &       &        &        &    &    &  &      &   &            Non - homogeneous mixing       \\
\hline 
\end{tabular}

\end{table}

\end{landscape}


\clearpage

\section*{Acknowledgments}
Funding for this study was provided by the Universidad de San Carlos de Guatemala for J.A.P., the Universidad del Norte for J.P.G. and the National Institutes of Health Grant 1R01GM117617 to J.M.P., R.D. Holt and J.K. Blackburn, 

\section*{Supplementary materials}
Materials and Methods\\
Figs. S1 to S14\\
References \textit{(33-46)}


\end{document}